\documentclass[10pt]{article}
\usepackage{graphicx}

\begin{document}

\title{Parallel GPU Implementation of \\ Iterative PCA Algorithms}

\author{M. Andrecut}

\date{ }

\maketitle

{\par\centering Institute for Biocomplexity and Informatics \par}
{\par\centering University of Calgary \par}
{\par\centering 2500 University Drive NW, Calgary \par}
{\par\centering Alberta, T2N 1N4, Canada \par}

\noindent

\begin{abstract}
Principal component analysis (PCA) is a key statistical technique for
multivariate data analysis. For large data sets the common approach to PCA
computation is based on the standard NIPALS-PCA algorithm, which
unfortunately suffers from loss of orthogonality, and therefore it's
applicability is usually limited to the estimation of the first few components.
Here we present an algorithm based on Gram-Schmidt
orthogonalization (called GS-PCA), which eliminates this
shortcoming of NIPALS-PCA. Also, we discuss the GPU (Graphics Processing
Unit) parallel implementation of both NIPALS-PCA and GS-PCA
algorithms. The numerical results show that the GPU parallel optimized
versions, based on CUBLAS (NVIDIA) are substantially faster (up to 12 times)
than the CPU optimized versions based on CBLAS (GNU Scientific Library).

\end{abstract}
\pagebreak

\section{Introduction}

Principal component analysis (PCA) is one of the most valuable results from
applied linear algebra, and probably the most popular method used for
compacting higher dimensional data sets into lower dimensional ones for data
analysis, visualization, feature extraction, or data compression \cite{Jackson01, Jolliffe01}. 
PCA provides a statistically optimal way
of dimensionality reduction by projecting the data onto a lower-dimensional
orthogonal subspace that captures as much of the variation of the data as
possible. Unfortunately, PCA quickly becomes quite expensive to compute for
high-dimensional data sets, where both the number of variables and samples
is high. Therefore, there is a real need in many applications to accelerate
the computation speed of PCA algorithms. For large data sets, the standard
approach is to use an iterative algorithm which computes the components
sequentially, and to avoid the global methods which calculate all the
components simultaneously. NIPALS-PCA \cite{Wold01} is the most frequently 
used iterative algorithm, and often considered as the standard PCA algorithm. 
However, for large data matrices, or matrices that have a high degree of column collinearity, 
NIPALS-PCA suffers from loss of
orthogonality, due to the errors accumulated in each iteration step \cite{Kramer01}.
Therefore, in practice it is only used to estimate the first few
components. Here, we address both the speed and orthogonality problems, and
we offer new solutions which eliminate these shortcomings of the iterative
PCA algorithms. 

We formulate an iterative PCA algorithm based on the Gram-Schmidt
re-orthogonalization, which we called GS-PCA. This algorithm is stable from
the orthogonality point of view, and if necessary, it can be used to
calculate the full set of principal components. The speed up issue is tackled with a
parallel implementation for Graphics Processing Units (GPUs). Here, we present  
the GPU parallel implementation of both NIPALS-PCA and GS-PCA algorithms. 
The numerical results show that the GPU parallel optimized versions, based on CUBLAS (NVIDIA) \cite{NVIDIA01}, 
are substantially faster (up to 12 times) than the CPU optimized versions based on CBLAS (GNU
Scientific Library) \cite{Galassi01}.

\section{Methods}
\subsection{Iterative Principal Component Analysis}
In the following description, the dataset to be analyzed is represented by
the $M\times N$ matrix $\mathbf{X}$. Each column, $\mathbf{X}^{(n)}$, $n=0,...,N-1$, 
contains all the observations of one attribute. Also, we
assume that each column is mean centered, i.e. if $\widetilde{\mathbf{X}}^{(n)}$ are the original vectors then 
\begin{equation} 
\mathbf{X}^{(n)}=\widetilde{\mathbf{X}}^{(n)}-N^{-1}\sum_{n=0}^{N-1}\widetilde{\mathbf{X}}^{(n)}\label{eq:01}
\end{equation}
PCA transforms the set of input column vectors $\mathbf{[X}^{(0)}|...|%
\mathbf{X}^{(N-1)}]$ into another set of column vectors $\mathbf{[T}%
^{(0)}|...|\mathbf{T}^{(N-1)}]$, called principal component scores.
This transformation has the property that most of the original data's
information content (or most of its variance) is stored in the first few component 
scores. This allows reduction of the data to a smaller number of dimensions,
with low information loss, simply by discarding the last component scores. Each component
is a linear combination of the original inputs and each component is orthogonal.
This linear transformation of the matrix $\mathbf{X}$ is specified by a $%
N\times N$ matrix $\mathbf{P}$ so that the matrix $\mathbf{X}$ is factorized
as: 
\begin{equation}
\mathbf{X}=\mathbf{TP}^{T}, \label{eq:02}
\end{equation}
 where $\mathbf{P}$ is known as the loadings
matrix.

There are several PCA algorithms in the literature, namely SVD (singular
value decomposition) and NIPALS (nonlinear iterative partial least squares),
which use the data matrix, and POWER and EVD (eigenvalue decomposition)
which use the covariance of the data matrix \cite{Jackson01, Jolliffe01}. SVD and EVD 
extract all the principal components simultaneously, while NIPALS and POWER calculate 
them sequentially. Unfortunately, the traditional implementation of PCA through
SVD or EVD quickly becomes prohibitive for very large data sets. In this case, 
an approximate solution can be more efficiently obtained using the iterative approach based on 
the NIPALS algorithm.

\subsection{NIPALS-PCA Algorithm }

The NIPALS-PCA algorithm can be described as following \cite{Wold01, Kramer01}. 
In the first step, the initial data $\mathbf{X}$ is copied into the residual matrix $\mathbf{R}$. 
Then, in the next steps the algorithm extracts iteratively one component at a time ($k=0, 1...,K\le N$) by repeated regressions of $\mathbf{X}^{T}$
on scores $\mathbf{T}^{(k)}$ to obtain improved loads $\mathbf{P}^{(k)}$, and
of $\mathbf{X}$ on these $\mathbf{P}^{(k)}$ to obtain improved scores $\mathbf{T}^{(k)}$. 
After the convergence is achieved, this process is followed by a deflation of the data matrix: 
\begin{equation}
\mathbf{X}\leftarrow \mathbf{X}-\mathbf{T}^{(k)}(\mathbf{P}^{(k)})^{T}. \label{eq:03}
\end{equation}
The convergence test consists in comparing two successive estimates of the
eigenvalue $\lambda $ and $\lambda ^{\prime }$. If the absolute difference 
$\left| \lambda ^{\prime }-\lambda \right| $ is smaller than some small error 
$\varepsilon $ then the convergence is achieved and the algorithm proceeds
to the deflation step. Using the NIPALS-PCA 
algorithm approach, the decomposition of the data matrix $\mathbf{X}$ takes the form: 
\begin{equation}
\mathbf{X}=\mathbf{T}_{(K)}\mathbf{P}_{(K)}^{T}+\mathbf{R}, \label{eq:04} 
\end{equation}
where $\mathbf{T}_{(K)}=\mathbf{[T}^{(0)}|...|\mathbf{T}^{(K-1)}]$ is the 
matrix formed using the first $K$ scores, 
$\mathbf{P}_{(K)}=\mathbf{[P}^{(0)}|...|\mathbf{P}^{(K-1)}]$ is the matrix of the first $K$ 
loadings, and $\mathbf{R}$ is the residual matrix.
The pseudo-code of the NIPALS-PCA algorithm is given below:

\bigskip

$\mathbf{R}\leftarrow \mathbf{X}$

for($k=0,...,K-1$) do

\qquad \{

\qquad $\lambda =0$

\qquad $\mathbf{T}^{(k)}\leftarrow \mathbf{R}^{(k)}$

\qquad for($j=0,...,J$) do

\qquad \qquad \{

\qquad \qquad $\mathbf{P}^{(k)}\leftarrow \mathbf{R}^{T}\mathbf{T}^{(k)}$

\qquad \qquad $\mathbf{P}^{(k)}\leftarrow \mathbf{P}^{(k)}\left\| \mathbf{P}%
^{(k)}\right\| ^{-1}$

\qquad \qquad $\mathbf{T}^{(k)}\leftarrow \mathbf{RP}^{(k)}$

\qquad \qquad $\lambda ^{\prime }\leftarrow \left\| \mathbf{T}^{(k)}\right\| $

\qquad \qquad if($\left| \lambda ^{\prime }-\lambda \right| \leq \varepsilon $) then break

\qquad \qquad $\lambda \leftarrow \lambda ^{\prime }$

\qquad \qquad \}

\qquad $\mathbf{R\leftarrow R-T}^{(k)}(\mathbf{P}^{(k)})^{T}$

\qquad \}

return $\mathbf{T}$, $\mathbf{P}$, $\mathbf{R}$ 

\bigskip

\subsection{GS-PCA Algorithm }

A well known shortcoming of the NIPALS-PCA algorithm is the loss of orthogonality \cite{Kramer01}. 
Both, the computed scores $\mathbf{T}^{(k)}$ and the loadings $\mathbf{P}^{(k)}$, are 
supposed to be orthogonal. However, because of the errors accumulated at each iteration step 
(which involves large matrix-vector operations) this 
orthogonality is quickly lost and in practice one can compute accurately only the 
first few components.
In order to stabilize the iterative PCA computation, from the orthogonality point of view, 
we propose an algorithm which is based on the Gram-Schmidt (GS) re-orthogonalization 
process. 
The classical GS algorithm (CGS) recursively constructs a set of orthonormal basis vectors 
for the subspace spanned by a given set of linearly independent normalized vectors \cite{Golub01}. 
It is well known that the CGS algorithm is also numerically unstable due to rounding errors. 
However, the CGS can be easily stabilized by a small modification obtaining the so called 
modified Gram-Schmidt (MGS) algorithm \cite{Bjorck01}. Unfortunately, the MGS algorithm cannot be expressed 
by Level-2 BLAS functions (matrix-vector operations) and therefore it requires a substantial 
amount of global communications, when implemented on a parallel computer \cite{Lingen01}. In contrast, 
the CGS algorithm can be easily expressed using matrix-vector operations and therefore 
it is more suitable for parallel implementation. Also, the numerical stability of CGS 
can be achieved by applying it iteratively \cite{Lingen01}.
In the proposed GS-PCA algorithm the re-orthogonalization correction is 
applied to both the scores and the loadings at each iteration step. 

For the pseudo-code formulation of the GS-PCA algorithm 
we prefer to use the truncated SVD description, since for $K=N$ the algorithm also returns the full SVD decomposition of the 
input matrix:
\begin{equation} 
\mathbf{X}=\mathbf{V}_{(K)}\mathbf{\Lambda }_{(K)}\mathbf{U}_{(K)}^{T}+\mathbf{R}\label{eq:05}
\end{equation}
where $\mathbf{V}_{(K)}$ and $\mathbf{U}_{(K)}$ are the first $K$ left and respectively rigth eigenvectors, $\mathbf{\Lambda}_{(K)}$ are the corresponding eigenvalues, 
and $\mathbf{R}$ is the residual. One can easily show that 
$\mathbf{T}_{(k)}=\mathbf{V}_{(K)}\mathbf{\Lambda }_{(K)}$ and $\mathbf{P}_{(K)}^{T}=\mathbf{U}_{(K)}^{T}$.
The pseudo-code of the GS-PCA algorithm can be formulated as following:

\bigskip

$\mathbf{R}\leftarrow \mathbf{X}$

for($k=0,...,K-1$) do

\qquad \{

$\qquad \mu =0$

$\qquad \mathbf{V}^{(k)}\leftarrow \mathbf{R}^{(k)}$

\qquad for($j=0,...,J$) do

\qquad \qquad \{

$\qquad \qquad \mathbf{U}^{(k)}\leftarrow \mathbf{R}^{T}\mathbf{V}^{(k)}$

\qquad \qquad if($k>0$) then

\qquad \qquad \qquad \{

$\qquad \qquad \qquad \mathbf{A\leftarrow U}_{(k)}^{T}\mathbf{U}^{(k)}$

$\qquad \qquad \qquad \mathbf{U}^{(k)}\leftarrow \mathbf{U}^{(k)}-\mathbf{U}_{(k)}\mathbf{A}$

\qquad \qquad \qquad \}

$\qquad \qquad \mathbf{U}^{(k)}\leftarrow \mathbf{U}^{(k)}\left\| \mathbf{U}^{(k)}\right\| ^{-1}$

$\qquad \qquad \mathbf{V}^{(k)}\leftarrow \mathbf{RU}^{(k)}$

\qquad \qquad if($k>0$) then

\qquad \qquad \qquad \{

\qquad \qquad \qquad $\mathbf{B\leftarrow V}_{(k)}^{T}\mathbf{V}^{(k)}$

$\qquad \qquad \qquad \mathbf{V}^{(k)}\leftarrow \mathbf{V}^{(k)}-\mathbf{V}_{(k)}\mathbf{B}$

\qquad \qquad \qquad \}

$\qquad \qquad \lambda _{k}\leftarrow \left\| \mathbf{V}^{(k)}\right\| $

\qquad \qquad $\mathbf{V}^{(k)}\leftarrow \mathbf{V}^{(k)}/\lambda _{k}$

\qquad \qquad if($\left| \lambda _{k}-\mu \right| \leq \varepsilon $) then break

$\qquad \qquad \mu \leftarrow \lambda _{k}$

\qquad \qquad \}

\qquad $\mathbf{R}\leftarrow \mathbf{R}-\lambda _{k}\mathbf{V}^{(k)}(\mathbf{U}^{(k)})^{T}$\qquad 

\qquad \}

$\mathbf{T}\leftarrow \mathbf{V\Lambda }$

$\mathbf{P}\leftarrow \mathbf{U}$

return $\mathbf{T}$, $\mathbf{P}$, $\mathbf{R}$ (for PCA) or $\mathbf{V}$, $\mathbf{U}$, $\mathbf{\Lambda}$ (for SVD)
\bigskip

\noindent One can see that in every iteration 
step, if $k>0$ then both the right (loads) and the left (scores) eigenvectors are re-orthonormalized. 
This procedure stabilizes the algorithm but it also increases a little bit the computational effort. However, this effort will   
be compensated by the efficiency of the parallel implementation. The GS-PCA algorithm assures the perfect orthogonality of 
both the loads and the scores. The errors accumulated in GS-PCA are only due to to the desired precision  $\varepsilon$ in the estimation 
of the eigenvalues $\lambda_{k}$. Also, for $K=N$ the GS-PCA algorithm returns a full SVD decomposition of 
the original matrix $\mathbf{X}$, with a maximum error $\varepsilon$ for eigenvalues and perfectly orthogonal left/right eigenvectors.

\section{Implementation Details}

The newly developed GPUs now include fully programmable processing units
that follow a stream programming model and support vectorized single and
double precision floating-point operations. For example, the CUDA computing
environment provides a standard C like language interface to the NVIDIA GPUs
\cite{NVIDIA01}. The computation is distributed into sequential grids, which are
organized as a set of thread blocks. The thread blocks are batches of
threads that execute together, sharing local memories and synchronizing at
specified barriers. An enormous number of blocks, each containing maximum
512 threads, can be launched in parallel in the grid. 

In our implementation of NIPALS-PCA and GS-PCA algorithms we use CUBLAS, a recent parallel implementation of
BLAS, developed by NVIDIA on top of the CUDA programming environment \cite{NVIDIA01}.
CUBLAS library provides functions for: 
\begin{itemize}

\item creating and destroying matrix and vector objects in GPU memory; 

\item transferring data from CPU mainmemory to GPU memory; 

\item executing BLAS on the GPU; 

\item transferring data from GPU memory back to the CPU mainmemory.

\end{itemize}

BLAS defines a set of low-level fundamental operations on vectors and matrices which
can be used to create optimized higher-level linear algebra functionality.
Highly efficient implementations of BLAS exist for most current
computer architectures and the specification of BLAS is widely adopted in
the development of high quality linear algebra software, such as the 
GNU Scientific Library (GSL) \cite{Galassi01}. We have selected GSL CBLAS, for
our host (CPU) implementation, due to its portability on various platforms
(Windows/Linux/OSX, Intel/AMD) and because it is free and easy to use in
combination with GCC (GNU Compiler). The GSL library provides a low-level
layer which corresponds directly to the C-language BLAS standard, referred
here as CBLAS, and a higher-level interface for operations on GSL vectors
and matrices.

The CBLAS (GNU Scientific Library) and respectively CUBLAS (NVIDIA
CUDA) implementations of the NIPALS-PCA and GS-PCA algorithms require the following
Level 1, 2, 3 BLAS functions (see the CBLAS/CUBLAS programming manuals for definition
details):\bigskip

\noindent CBLAS (Level 2): gsl\_blas\_dgemv

\noindent CUBLAS (Level 2): cublasDgemv

\begin{itemize}
\item  computes in double precision the matrix-vector product and sum: 
\begin{equation}
\mathbf{y\leftarrow }\alpha \mathbf{Ax}+\beta \quad or\quad \mathbf{%
y\leftarrow }\alpha \mathbf{A}^{T}\mathbf{x}+\beta 
\end{equation}
$\alpha $ and $\beta $ are double precision scalars, and $\mathbf{x}$ and $%
\mathbf{y}$ are double precision vectors. $\mathbf{A}$ is a matrix
consisting of double precision elements. Matrix $\mathbf{A}$ is stored in
column major format.
\end{itemize}

\noindent \noindent CBLAS (Level 1): gsl\_blas\_daxpy

\noindent CUBLAS (Level 1): cublasDaxpy

\begin{itemize}
\item  computes the double precision sum: 
\begin{equation}
\mathbf{y}\leftarrow \alpha \mathbf{x}+\mathbf{y},
\end{equation}
multiplies double precision vector $\mathbf{x}$ by double precision scalar $%
\alpha $ and adds the result to double precision vector $\mathbf{y}$.
\end{itemize}

\noindent CBLAS (Level 1): gsl\_blas\_dnrm2

\noindent CUBLAS (Level 1): cublasDnrm2

\begin{itemize}
\item  computes the Euclidean norm of a double precision vector $%
\mathbf{x}$: 
\begin{equation}
\left\| \mathbf{x}\right\| _{2}\leftarrow \sqrt{\sum_{m=0}^{M}x_{m}^{2}}.
\end{equation}
\end{itemize}

\noindent CBLAS (Level 3): gsl\_blas\_dger

\noindent CUBLAS (Level 3): cublasDger

\begin{itemize}
\item  computes in double precision the matrix-matrix sum: 
\begin{equation}
\mathbf{A\leftarrow }\alpha \mathbf{xy}^{T}+\mathbf{A,}
\end{equation}
where $\alpha $ is a double precision scalar, $\mathbf{x}$ is an $M$ element
double precision vector, $\mathbf{y}$ is an $N$ element double precision
vector, and $\mathbf{A}$ is an $M\times N$ matrix consisting of double
precision elements. Matrix $\mathbf{A}$ is stored in column major format.
\end{itemize}

\noindent These are the critical functions/kernels which are efficiently
exploited in the parallel CUBLAS implementation. The other involved
functions are for vector/matrix memory allocation and vector/matrix
accessing, device (GPU) initialization, host-device data transfer and error
handling. 

In the CUBLAS implementation, the data
space is allocated both on host (CPU) mainmemory and on device (GPU) memory.
After the data is initialized on host it is transferred on device, where the
main parallel computation occurs. The results are then transferred back on
host memory. 

The double precision code for CBLAS and CUBLAS implementations are given in the Appendices 1, 2, 3 and 4. 
These implementations can be easily modified in
order to meet the end user's specifications. For example the single precision BLAS data
allocation and vector/matrix accessing functions have the prefix
gsl\_vector\_float, gsl\_matrix\_float etc. (see CBLAS/CUBLAS programming
manuals for details). For convenience, we have included a timer which measures
the time of all main operations involved by the algorithm. 

The CBLAS (GSL) versions can be compiled using the following commands:

\begin{verbatim}
              g++ -O2 nipals_pca.cpp -lgsl -lgslcblas -lm
\end{verbatim}
\begin{verbatim}
              g++ -O2 gs_pca.cpp -lgsl -lgslcblas -lm
\end{verbatim}

\smallskip

An obvious requirement for the CUBLAS (NVIDIA) version is the presence of 
an NVIDIA GPU installed. The GPU must be minimum GTX260, GTX280 or Tesla C1060, otherwise there is no support for 
double precision. Also, the NVIDIA Driver, Toolkit and SDK must be correctly instaled. 
In order to compile the CUBLAS (NVIDIA) version we recommend to use the following 
simple make file:

\bigskip

\begin{verbatim}
#######################################################
# Build script for the CUBLAS (NVIDIA) implementation #
# of the NIPALS-PCA and GS-PCA project                #
#######################################################

# Add source files here 
# (comment/uncomment your choice)

  EXECUTABLE      := nipals_pca  # for NIPALS-PCA
# EXECUTABLE      := gs_pca      # for GS-PCA

# C/C++ source files (compiled with gcc / c++)
# (comment/uncomment your choice)

  CFILES          := nipals_pca.c # for NIPALS-PCA
# CFILES          := gs_pca.c     # for GS-PCA 

# Additional libraries needed by the project

  USECUBLAS       := 1

# Rules and targets
  include ../../common/common.mk

#######################################################
\end{verbatim}

\section{Results and Conclusion}

The numerical tests have been carried out on the following system: 
AMD Phenom 9950 CPU (2.6GHz); XFX GTX280 GPU; NVIDIA Linux
64-bit driver (177.67); CUDA Toolkit and SDK 2.0; Ubuntu Linux 64-bit
8.04.1, GNU Scientific Library v.1.11; Compilers: GCC (GNU), NVCC (NVIDIA).
The GPU used is a high end graphics card
solution with 240 stream processors and 1Gb DDR3 RAM, which supports both
single and double precision and it is theoretically capable of 1 Tflop
computational power.

In Figure 1 we give the CPU vs GPU execution time as a function of the size of the randomly generated input 
matrix $\mathbf{X}$ ($M\in [5\times 10^{2}, 1.5\times 10^{4}], N=M/2, K=10, \varepsilon =10^{-7}$). 
The time gap between CPU and GPU increases very fast by
increasing the size of the input matrix, and the CPU time versus the GPU time
reaches a maximum for $M=1.5\times 10^{4}$, where the GPU is about 12 times
faster than the CPU. The GS-PCA algorithm is
only about 5-7\% slower than the standard NIPLALS-PCA algorithm, in both CPU
and GPU implementation. These results also show that the GPU performance is
dependent on the scale of the problem. Thus, in order to exploit efficiently
the massive parallelism of GPUs and to effectively use the hardware
capabilities, the problem itself needs to scale accordingly, such that
thousands of threads are defined and used in computation.

\begin{figure}[!htp]
\centerline{\includegraphics{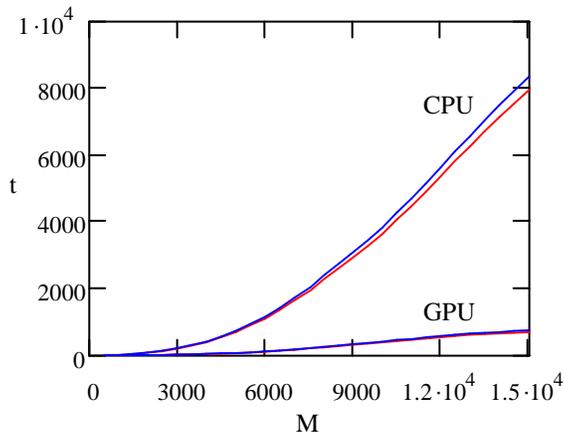}}
\caption{GPU vs CPU execution time of the NIPALS-PCA and GS-PCA algorithms
as a function of the size of the input matrix $\mathbf{X}$ ($M, N=M/2, K=10$%
, Red = NIPALS-PCA, Blue = GS-PCA).}
\label{fig:01}
\end{figure}

In conclusion, we have presented an iterative GS-PCA algorithm based on Gram-Schmidt
re-orthogonalization. The GS-PCA algorithm assures the perfect orthogonality of 
both the loads and the scores, and thus totally eliminates the loss of orthogonality present 
in the standard NIPALS-PCA algorithm. Also, we have discussed the GPU parallel implementation of both NIPALS-PCA and GS-PCA
algorithms. We have shown that the GPU parallel optimized versions, based on CUBLAS (NVIDIA) are substantially faster (up to 12 times)
than the CPU optimized versions based on CBLAS (GNU Scientific Library).

\section{Acknowledgement}

The author acknowledge the financial support from IBI and the University of Calgary.

\pagebreak

\section*{Appendix 1: nipals\_pca.cpp}

\begin{verbatim}

// C/C++ example for the CBLAS (GNU Scientific Library) 
// implementation of the NIPALS-PCA algorithm
// M. Andrecut (c) 2008
// 
// Compile with: g++ -O3 nipals.cpp -lgsl -lgslcblas -lm

// includes, system 

#include <math.h>
#include <time.h>

// includes, GSL & CBLAS

#include <gsl/gsl_vector.h>
#include <gsl/gsl_matrix.h>
#include <gsl/gsl_blas.h>

// declarations

int nipals_gsl(int, int, int, gsl_matrix *, 
               gsl_matrix *, gsl_matrix *);

int print_results(int, int, int, 
                  gsl_matrix *, gsl_matrix *, 
                  gsl_matrix *, gsl_matrix *);

// main
int main(int argc, char** argv)
{
// PCA model: X = TP' + R

// input: X, MxN matrix (data)
// input: M = number of rows in X
// input: N = number of columns in X 
// input: K = number of components (K<=N)

// output: T, MxK scores matrix 
// output: P, NxN loads matrix
// output: R, MxN residual matrix

int M = 1000, m; 
int N = M/2, n;
int K = 25;

printf("\nProblem dimensions: MxN=%dx%d, K=%d\n", M, N, K);

// initialize srand and clock 

srand(time(NULL));

clock_t start=clock(); 

double htime;
	
// initiallize some random test data X

gsl_matrix *X = gsl_matrix_alloc(M, N);
for(m=0; m<M; m++)  
 {
 for(n=0; n<N; n++)  
  {
  gsl_matrix_set(X, m, n, rand()/(double)RAND_MAX);
  }
 }
	
// allocate memory for T, P, R

gsl_matrix *T = gsl_matrix_alloc(M, K);

gsl_matrix *P = gsl_matrix_alloc(N, K);

gsl_matrix *R = gsl_matrix_alloc(M, N);

htime = ((double)clock()-start)/CLOCKS_PER_SEC;

printf("\nTime for data allocation: %f\n", htime); 

// call the nipals_gsl() function

start=clock();

gsl_matrix_memcpy(R, X);  

nipals_gsl(M, N, K, T, P, R);

htime = ((double)clock()-start)/CLOCKS_PER_SEC;

printf("\n\nTime for NIPALS-PCA computation on host: %f\n", htime); 

// the results are in T, P, R

print_results(M, N, K, X, T, P, R);

// memory clean up and shutdown

gsl_matrix_free(R);

gsl_matrix_free(P);

gsl_matrix_free(T);

gsl_matrix_free(X);
    
printf("\nPress ENTER to exit...\n");  getchar();

return EXIT_SUCCESS;
}

int nipals_gsl(int M, int N, int K, gsl_matrix *T, 
               gsl_matrix *P, gsl_matrix *R)
{
// PCA model: X = TP' + R 

// input: X, MxN matrix (data)
// input: M = number of rows in X
// input: N = number of columns in X
// input: K = number of components (K<=N)

// output: T, MxK scores matrix 
// output: P, NxK loads matrix
// output: R, MxN residual matrix (X is initially copied in R)
	
// maximum number of iterations

int J = 10000;
	
// max error

double er = 1.0e-7;
	
// some useful pointers

double *a = (double*)calloc(1, sizeof(a)); 

double *b = (double*)calloc(1, sizeof(b)); 
	
int k, n, j; 

// mean center the data 

gsl_vector *U = gsl_vector_calloc(M);

for(n=0; n<N; n++) 
 {
 gsl_blas_daxpy(1.0, &gsl_matrix_column(R, n).vector, U);	
 }

for(n=0; n<N; n++) 
 {
 gsl_blas_daxpy(-1.0/N, U, &gsl_matrix_column(R, n).vector);
 }
 
for(k=0; k<K; k++)
 {
 gsl_blas_dcopy(&gsl_matrix_column(R, k).vector, 
                &gsl_matrix_column(T, k).vector);

 *a = 0.0;

 for(j=0; j<J; j++)
  {
  gsl_blas_dgemv(CblasTrans, 1.0, R, 
                 &gsl_matrix_column(T, k).vector, 
                 0.0, &gsl_matrix_column(P, k).vector);	

  gsl_blas_dscal(1.0/gsl_blas_dnrm2(&gsl_matrix_column(P, k).vector), 
                 &gsl_matrix_column(P, k).vector);

  gsl_blas_dgemv(CblasNoTrans, 1.0, R, 
                 &gsl_matrix_column(P, k).vector, 
                 0.0, &gsl_matrix_column(T, k).vector);				

  *b = gsl_blas_dnrm2(&gsl_matrix_column(T, k).vector); 	

  if(fabs(*a - *b) < er*(*b)) break;

  *a = *b;
  }
 gsl_blas_dger(-1.0, &gsl_matrix_column(T, k).vector, 
                &gsl_matrix_column(P, k).vector, R);
 }
		
// clean up memory

free(a); 

free(b);

gsl_vector_free(U);

return EXIT_SUCCESS;
}


int print_results(int M, int N, int K, 
                  gsl_matrix *X, gsl_matrix *T, 
                  gsl_matrix *P, gsl_matrix *R)
{
int m, n;

// If M < 13 print the results on screen 

if(M > 12) return EXIT_SUCCESS;

printf("\nX\n");

for(m=0; m<M; m++)
 {
 for(n=0; n<N; n++)
  {
  printf("%+f  ", gsl_matrix_get(X, m, n));
  }
  printf("\n");
 }

printf("\nT\n");

for(m=0; m<M; m++)
 {
 for(n=0; n<K; n++)
  {
  printf("%+f  ", gsl_matrix_get(T, m, n));
  }
  printf("\n");
 }		

gsl_matrix *F = gsl_matrix_alloc(K, K);		

gsl_blas_dgemm (CblasTrans, CblasNoTrans, 1.0, T, T, 0.0, F);

printf("\nT' * T\n");

for(m=0; m<K; m++)
 {
 for(n=0; n<K; n++)
  {
  printf("%+f  ", gsl_matrix_get(F, m, n));
  }
  printf("\n");
 }

gsl_matrix_free(F);	

printf("\nP\n");

for(m=0; m<N; m++)
 {
 for(n=0; n<K; n++)
  {
  printf("%+f  ", gsl_matrix_get(P, m, n));
  }
 printf("\n");
 }

gsl_matrix *G = gsl_matrix_alloc(K, K);		

gsl_blas_dgemm (CblasTrans, CblasNoTrans, 1.0, P, P, 0.0, G);

printf("\nP' * P\n");

for(m=0; m<K; m++)
 {
 for(n=0; n<K; n++)
  {
  printf("%+f  ", gsl_matrix_get(G, m, n));
  }
 printf("\n");
 }

gsl_matrix_free(G);

printf("\nR\n");

for(m=0; m<M; m++)
 {
 for(n=0; n<N; n++)
  {
  printf("%+f  ", gsl_matrix_get(R, m, n));
  }
 printf("\n");
 }

return EXIT_SUCCESS;
}
\end{verbatim}

\pagebreak

\section*{Appendix 2: nipals\_pca.c}
\begin{verbatim}
// C/C++ example for the CUBLAS (NVIDIA) 
// implementation of NIPALS-PCA algorithm
//
// M. Andrecut (c) 2008


// includes, system 

#include <stdio.h>
#include <stdlib.h>
#include <string.h>
#include <time.h>

// includes, cuda 

#include <cublas.h>

// matrix indexing convention 

#define id(m, n, ld) (((n) * (ld) + (m)))

// declarations 

int nipals_cublas(int, int, int, 
                  double *, double *, 
                  double *);

int print_results(int, int, int, 
                  double *, double *, 
                  double *, double *);

// main 
int main(int argc, char** argv)
{ 
// PCA model: X = T * P' + R

// input: X, MxN matrix (data) 
// input: M = number of rows in X
// input: N = number of columns in X 
// input: K = number of components (K<=N)

// output: T, MxK scores matrix 
// output: P, NxN loads matrix
// output: R, MxN residual matrix
	
int M = 1000, m;
int N = M/2, n;
int K = 25;
		
printf("\nProblem dimensions: MxN=%dx%d, K=%d\n", M, N, K);

// initialize srand and clock 

srand(time(NULL));

clock_t start=clock(); 

double dtime;
	
// initialize cublas

cublasStatus status;

status = cublasInit();

if(status != CUBLAS_STATUS_SUCCESS) 
 {
 fprintf(stderr, "! CUBLAS initialization error\n");  
 return EXIT_FAILURE;
 }

// initiallize some random test data X

double *X; 

X = (double*)malloc(M*N * sizeof(X[0]));

if(X == 0) 
 {
 fprintf(stderr, "! host memory allocation error: X\n");  
 return EXIT_FAILURE;
 }

for(m = 0; m < M; m++) 
 {
 for(n = 0; n < N; n++) 
  {
  X[id(m, n, M)] = rand() / (double)RAND_MAX; 
  }
 }

// allocate host memory for T, P, R

double *T; 

T = (double*)malloc(M*K * sizeof(T[0]));;

if(T == 0)
 {
 fprintf(stderr, "! host memory allocation error: T\n");  
 return EXIT_FAILURE;
 }

double *P; 

P = (double*)malloc(N*K * sizeof(P[0]));;

if(P == 0)
 {
 fprintf(stderr, "! host memory allocation error: P\n");  
 return EXIT_FAILURE;
 }

double *R; 

R = (double*)malloc(M*N * sizeof(R[0]));;

if(R == 0)
 {
 fprintf(stderr, "! host memory allocation error: R\n");  
 return EXIT_FAILURE;
 }
			
dtime = ((double)clock() - start)/CLOCKS_PER_SEC;

printf("\nTime for data allocation: %f\n", dtime);   
	
// call nipals_cublas() 

start=clock(); 

memcpy(R, X, M*N * sizeof(X[0]));

nipals_cublas(M, N, K, T, P, R);

dtime = ((double)clock() - start)/CLOCKS_PER_SEC;

printf("\nTime for NIPALS-PCA computation on device: %f\n", dtime); 

print_results(M, N, K, X, T, P, R);

// memory clean up 

free(R); 

free(P); 	

free(T);  

free(X);

// shutdown

status = cublasShutdown();

if(status != CUBLAS_STATUS_SUCCESS)  
 {
 fprintf(stderr, "! cublas shutdown error\n"); 
 return EXIT_FAILURE;
 }

if(argc <= 1 || strcmp(argv[1], "-noprompt")) 
 {
 printf("\nPress ENTER to exit...\n");  
 getchar();
 }

return EXIT_SUCCESS;
}


int nipals_cublas(int M, int N, int K, 
                  double *T, double *P, 
                  double *R)
{
// PCA model: X = T * P' + R

// input: X, MxN matrix (data)
// input: M = number of rows in X
// input: N = number of columns in X (N<=M)
// input: K = number of components (K<N)

// output: T, MxK scores matrix 
// output: P, NxK loads matrix
// output: R, MxN residual matrix

// CUBLAS error handling

cublasStatus status;

// maximum number of iterations 

int J = 10000;

// max error

double er = 1.0e-7; 

int k, n, j;
	
// transfer the host matrix X to device matrix dR 

double *dR = 0; 

status = cublasAlloc(M*N, sizeof(dR[0]), (void**)&dR);

if(status != CUBLAS_STATUS_SUCCESS)
 {
 fprintf (stderr, "! device memory allocation error (dR)\n"); 
 return EXIT_FAILURE;
 }    

status = cublasSetMatrix(M, N, sizeof(R[0]), R, M, dR, M);

if(status != CUBLAS_STATUS_SUCCESS) 
 {
 fprintf(stderr, "! device access error (write dR)\n"); 
 return EXIT_FAILURE;
 }

// allocate device memory for T, P

double *dT = 0; 

status = cublasAlloc(M*K, sizeof(dT[0]), (void**)&dT);

if(status != CUBLAS_STATUS_SUCCESS)
 {
 fprintf(stderr, "! device memory allocation error (dT)\n"); 
 return EXIT_FAILURE;
 }  
  
double *dP = 0; 

status = cublasAlloc(N*K, sizeof(dP[0]), (void**)&dP);

if(status != CUBLAS_STATUS_SUCCESS)
 {
 fprintf(stderr, "! device memory allocation error (dP)\n"); 
 return EXIT_FAILURE;
 }    

// mean center the data 

double *dU = 0; 

status = cublasAlloc(M, sizeof(dU[0]), (void**)&dU);

if(status != CUBLAS_STATUS_SUCCESS) 
 {
 fprintf(stderr, "! device memory allocation error (dU)\n"); 
 return EXIT_FAILURE;
 }  

cublasDcopy(M, &dR[0], 1, dU, 1);   

for(n=1; n<N; n++) 
 {
 cublasDaxpy(M, 1.0, &dR[n*M], 1, dU, 1);
 }

for(n=0; n<N; n++) 
 {
 cublasDaxpy(M, -1.0/N, dU, 1, &dR[n*M], 1);	
 }
    
double a, b; 
		
for(k=0; k<K; k++)
 {
 cublasDcopy(M, &dR[k*M], 1, &dT[k*M], 1);

 a = 0.0; 

 for(j=0; j<J; j++)
  {
  cublasDgemv('t', M, N, 1.0, dR, M, &dT[k*M], 1, 0.0, &dP[k*N], 1);

  cublasDscal(N, 1.0/cublasDnrm2(N, &dP[k*N], 1), &dP[k*N], 1);

  cublasDgemv('n', M, N, 1.0, dR, M, &dP[k*N], 1, 0.0, &dT[k*M], 1);

  b = cublasDnrm2(M, &dT[k*M], 1);

  if(fabs(a - b) < er*b) break;

  a = b;
  }
 cublasDger(M, N, -1.0, &dT[k*M], 1, &dP[k*N], 1, dR, M);
 }		

// transfer device dT to host T

cublasGetMatrix(M, K, sizeof(dT[0]), dT, M, T, M);

// transfer device dP to host P

cublasGetMatrix(N, K, sizeof(dP[0]), dP, N, P, N);

// transfer device dR to host R

cublasGetMatrix(M, N, sizeof(dR[0]), dR, M, R, M);
	
// clean up memory 

status = cublasFree(dP);

status = cublasFree(dT);

status = cublasFree(dR);

return EXIT_SUCCESS;
}


int print_results(int M, int N, int K, 
                  double *X, double *T, 
                  double *P, double *R)
{
int m, n, k;
	
// If M < 13 print the results on screen

if(M > 12) return EXIT_SUCCESS;

printf("\nX\n");

for(m=0; m<M; m++)
 {
 for(n=0; n<N; n++)
  {
  printf("%+f  ", X[id( m, n,M)]);
  }
  printf("\n");
 }

printf("\nT\n");

for(m=0; m<M; m++)
 {
 for(n=0; n<K; n++)
  {
  printf("%+f  ", T[id(m, n, M)]);
  }
 printf("\n");
 }

double a;

printf("\nT' * T\n");

for(m = 0; m<K; m++)
 {
 for(n=0; n<K; n++)
  {
  a=0; 
  for(k=0; k<M; k++) 
   {
   a = a + T[id(k, m, M)] * T[id(k, n, M)];
   }
  printf("%+f  ", a);
  }
 printf("\n");
 }		

printf("\nP\n");

for(m=0; m<N; m++)
 {
 for(n=0; n<K; n++)
  {
  printf("%+f  ", P[id(m, n, N)]);
  }
 printf("\n");
 }

printf("\nP' * P\n");

for(m = 0; m<K; m++)
 {
 for(n=0; n<K; n++)
  {
  a=0; 
  for(k=0; k<N; k++) 
   {
   a = a + P[id(k, m, N)] * P[id(k, n, N)];
   }
  printf("%+f  ", a);
  }
 printf("\n");
 }

printf("\nR\n");

for(m=0; m<M; m++)
 {
 for(n=0; n<N; n++)
  {
  printf("%+f  ", R[id( m, n,M)]);
  }
 printf("\n");
 }

return EXIT_SUCCESS;
}
\end{verbatim}

\pagebreak

\section*{Appendix 3: gs\_pca.cpp}
\begin{verbatim}

// C/C++ example for the CBLAS (GNU Scientific Library) 
// implementation of PCA-GS algorithm
//
// M. Andrecut (c) 2008

// includes, system 

#include <math.h>

#include <time.h>

// includes, GSL & CBLAS

#include <gsl/gsl_vector.h>

#include <gsl/gsl_matrix.h>

#include <gsl/gsl_blas.h>

// declarations

int gs_pca_gsl(int, int, int, 
               gsl_matrix *, gsl_matrix *, 
               gsl_matrix *);

int print_results(int, int, int, 
                  gsl_matrix *, gsl_matrix *, 
                  gsl_matrix *, gsl_matrix *);

// main
int main(int argc, char** argv)
{
// PCA model: X = TP' + R

// input: X, MxN matrix (data)
// input: M = number of rows in X
// input: N = number of columns in X 
// input: K = number of components (K<=N)

// output: T, MxK scores matrix 
// output: P, NxK loads matrix
// output: R, MxN residual matrix

int M = 1000, m; 
int N = M/2, n;
int K = 10;
	
printf("\nProblem dimensions: MxN=%dx%d, K=%d", M, N, K);

// initialize srand and clock 

srand(time(NULL));

clock_t start=clock(); 

double htime;
	
// initiallize some random test data X

gsl_matrix *X = gsl_matrix_alloc (M, N);

for(m=0; m<M; m++)  
 {
 for(n=0; n<N; n++)  
  {
  gsl_matrix_set(X, m, n, rand()/(double)RAND_MAX);
  }
 }
	
// allocate memory for T, P, R

gsl_matrix *T = gsl_matrix_alloc (M, K);

gsl_matrix *P = gsl_matrix_alloc (N, K);

gsl_matrix *R = gsl_matrix_alloc (M, N);

htime = ((double)clock()-start)/CLOCKS_PER_SEC;

printf("\nTime for data allocation: %f\n", htime); 

// call gs_pca_gsl

start=clock();

gsl_matrix_memcpy (R, X);  

gs_pca_gsl(M, N, K, T, P, R);

htime = ((double)clock()-start)/CLOCKS_PER_SEC;

printf("\n\nTime for GS-PCA host computation: %f\n", htime); 

// the results are in T, P, R

print_results(M, N, K, X, T, P, R);

// memory clean up and shutdown

gsl_matrix_free(R);

gsl_matrix_free(P);

gsl_matrix_free(T);

gsl_matrix_free(X);
    
printf("\nPress ENTER to exit...\n");  getchar();

return EXIT_SUCCESS;
}


int gs_pca_gsl(int M, int N, int K, 
               gsl_matrix *T, gsl_matrix *P, 
               gsl_matrix *R)
{
// PCA model: X = TLP' + R

// input: X, MxN matrix (data, copied in R)
// input: M = number of rows in X
// input: N = number of columns in X
// input: K = number of components (K<=N)

// output: T, MxK left eigenvectors 
// output: P, NxK right eigenvectors
// output: L, Kx1 eigenvalues
// output: R, MxN residual
	
// maximum number of iterations

int J = 10000;
	
// max error

double er = 1.0e-7;
double a;

int n, k, j; 

// mean center the data 

gsl_vector *U = gsl_vector_calloc (M);

for(n=0; n<N; n++) 
 {
 gsl_blas_daxpy(1.0, &gsl_matrix_column(R, n).vector, U);	
 }

for(n=0; n<N; n++) 
 {
 gsl_blas_daxpy(-1.0/N, U, &gsl_matrix_column(R, n).vector);
 }

// allocate memory fo eigenvalues

gsl_vector *L = gsl_vector_alloc(K);
	
// gs_pca

for(k=0; k<K; k++)
 {
 gsl_blas_dcopy(&gsl_matrix_column(R, k).vector, 
                &gsl_matrix_column(T, k).vector);

 a = 0.0;

 for(j=0; j<J; j++)
  {
  gsl_blas_dgemv(CblasTrans, 1.0, R, 
                 &gsl_matrix_column(T, k).vector, 
                 0.0, &gsl_matrix_column(P, k).vector);	

  if(k>0)
   {
   gsl_blas_dgemv(CblasTrans, 1.0, 
                  &gsl_matrix_submatrix (P, 0, 0, N, k).matrix, 
                  &gsl_matrix_column(P, k).vector, 0.0, 
                  &gsl_vector_subvector (U, 0, k).vector);

   gsl_blas_dgemv(CblasNoTrans, -1.0, 
                  &gsl_matrix_submatrix (P, 0, 0, N, k).matrix, 
                  &gsl_vector_subvector (U, 0, k).vector, 1.0, 
                  &gsl_matrix_column(P, k).vector);					
   }

  gsl_blas_dscal(1.0/gsl_blas_dnrm2(&gsl_matrix_column(P, k).vector), 
                 &gsl_matrix_column(P, k).vector);

  gsl_blas_dgemv(CblasNoTrans, 1.0, R, &gsl_matrix_column(P, k).vector, 
                 0.0, &gsl_matrix_column(T, k).vector);		

  if(k>0)
   {
   gsl_blas_dgemv(CblasTrans, 1.0, 
                  &gsl_matrix_submatrix (T, 0, 0, M, k).matrix, 
                  &gsl_matrix_column(T, k).vector, 0.0, 
                  &gsl_vector_subvector (U, 0, k).vector);

   gsl_blas_dgemv(CblasNoTrans, -1.0, 
                  &gsl_matrix_submatrix (T, 0, 0, M, k).matrix, 
                  &gsl_vector_subvector (U, 0, k).vector, 1.0, 
                  &gsl_matrix_column(T, k).vector);					
   }

  gsl_vector_set(L, k, gsl_blas_dnrm2(&gsl_matrix_column(T, k).vector));		

  gsl_blas_dscal(1.0/gsl_vector_get(L, k), 
                 &gsl_matrix_column(T, k).vector);

  if(fabs(a - gsl_vector_get(L, k)) < er*gsl_vector_get(L, k)) break;

  a = gsl_vector_get(L, k);
  }
  gsl_blas_dger (-gsl_vector_get(L, k), 
                 &gsl_matrix_column(T, k).vector, 
                 &gsl_matrix_column(P, k).vector, R);
 }		

for(k=0; k<K; k++)
 {
 gsl_blas_dscal (gsl_vector_get(L, k), 
                 &gsl_matrix_column(T, k).vector);
 }

// memory clean up

gsl_vector_free(L);

gsl_vector_free(U);

return EXIT_SUCCESS;
}


int print_results(int M, int N, int K, 
                  gsl_matrix *X, gsl_matrix *T, 
                  gsl_matrix *P, gsl_matrix *R)
{
	int m, n;

/* If M < 13 print the results on screen */	
	
if(M > 12) return EXIT_SUCCESS;

printf("\nX\n");

for(m=0; m<M; m++)
 {
 for(n=0; n<N; n++)
  {
  printf("%+f  ", gsl_matrix_get(X, m, n));
  }
 printf("\n");
 }

printf("\nT\n");

for(m=0; m<M; m++)
 {
 for(n=0; n<K; n++)
  {
  printf("%+f  ", gsl_matrix_get(T, m, n));
  }
 printf("\n");
 }		

gsl_matrix *F = gsl_matrix_alloc (K, K);		

gsl_blas_dgemm(CblasTrans, CblasNoTrans, 1.0, T, T, 0.0, F);

printf("\nT' * T\n");

for(m=0; m<K; m++)
 {
 for(n=0; n<K; n++)
  {
  printf("%+f  ", gsl_matrix_get(F, m, n));
  }
 printf("\n");
 }

gsl_matrix_free(F);	

printf("\nP\n");

for(m=0; m<N; m++)
 {
 for(n=0; n<K; n++)
  {
  printf("%+f  ", gsl_matrix_get(P, m, n));
  }
 printf("\n");
 }

gsl_matrix *G = gsl_matrix_alloc (K, K);		

gsl_blas_dgemm (CblasTrans, CblasNoTrans, 1.0, P, P, 0.0, G);

printf("\nP' * P\n");

for(m=0; m<K; m++)
 {
 for(n=0; n<K; n++)
  {
  printf("%+f  ", gsl_matrix_get(G, m, n));
  }
 printf("\n");
 }

gsl_matrix_free(G);

printf("\nR\n");

for(m=0; m<M; m++)
 {
 for(n=0; n<N; n++)
  {
  printf("%+f  ", gsl_matrix_get(R, m, n));
  }
 printf("\n");
 }

return EXIT_SUCCESS;
}



\end{verbatim}

\pagebreak

\section*{Appendix 4: gs\_pca.c}
\begin{verbatim}

// C/C++ example for the CUBLAS (NVIDIA) 
// implementation of PCA-GS algorithm
//
// M. Andrecut (c) 2008

// includes, system 

#include <stdio.h>
#include <stdlib.h>
#include <string.h>
#include <time.h>

// includes, cuda 

#include <cublas.h>

// matrix indexing convention 
#define id(m, n, ld) (((n) * (ld) + (m)))

// declarations 

int gs_pca_cublas(int, int, int, double *, double *, double *);

int print_results(int, int, int, double *, double *, double *, double *);

// main 
int main(int argc, char** argv)
{ 
// PCA model: X = TP' + R

// input: X, MxN matrix (data)
// input: M = number of rows in X
// input: N = number of columns in X 
// input: K = number of components (K<=N)

// output: T, MxK scores matrix 
// output: P, NxK loads matrix
// output: R, MxN residual matrix

int M = 1000, m; 
int N = M/2, n;
int K = 10;
	
printf("\nProblem dimensions: MxN=%dx%d, K=%d", M, N, K);

// initialize srand and clock 

srand (time(NULL));

clock_t start=clock(); 

double dtime;
	
// initialize cublas

cublasStatus status;

status = cublasInit();

if(status != CUBLAS_STATUS_SUCCESS) 
 {
 fprintf(stderr, "! CUBLAS initialization error\n");  
 return EXIT_FAILURE;
 }

// initiallize some random test data X

double *X; 

X = (double*)malloc(M*N * sizeof(X[0]));

if(X == 0) 
 {
 fprintf (stderr, "! host memory allocation error: X\n");  
 return EXIT_FAILURE;
 }

for(m = 0; m < M; m++) 
 {
 for(n = 0; n < N; n++) 
  {
  X[id(m, n, M)] = rand() / (double)RAND_MAX; 
  }
 }

// allocate host memory for T, P, R

double *T; 

T = (double*)malloc(M*K * sizeof(T[0]));;
if(T == 0)
 {
 fprintf(stderr, "! host memory allocation error: T\n");  
 return EXIT_FAILURE;
 }

double *P; 

P = (double*)malloc(N*K * sizeof(P[0]));;

if(P == 0)
 {
 fprintf(stderr, "! host memory allocation error: P\n");  
 return EXIT_FAILURE;
 }

double *R; 

R = (double*)malloc(M*N * sizeof(R[0]));;

if(R == 0)
 {
 fprintf(stderr, "! host memory allocation error: R\n");  
 return EXIT_FAILURE;
 }
			
dtime = ((double)clock()-start)/CLOCKS_PER_SEC;

printf("\nTime for data allocation: %f\n", dtime);       

//  call gs_pca_cublas

start=clock();

memcpy(R, X, M*N * sizeof(X[0]));

gs_pca_cublas(M, N, K, T, P, R);

dtime = ((double)clock()-start)/CLOCKS_PER_SEC;

printf("\nTime for device GS-PCA computation: %f\n", dtime); 

// the results are in T, P, R

print_results(M, N, K, X, T, P, R);

// clean up memory

free(P); 	

free(T);  

free(X);

// shutdown 

status = cublasShutdown();
if(status != CUBLAS_STATUS_SUCCESS)  
 {
 fprintf (stderr, "! cublas shutdown error\n"); 
 return EXIT_FAILURE;
 }
 
if(argc <= 1 || strcmp(argv[1], "-noprompt")) 
 {
 printf("\nPress ENTER to exit...\n");  getchar();
 }
 
return EXIT_SUCCESS;
}


int gs_pca_cublas(int M, int N, int K, 
                  double *T, double *P, 
                  double *R)
{
// PCA model: X = TP' + R

// input: X, MxN matrix (data)
// input: M = number of rows in X
// input: N = number of columns in X
// input: K = number of components (K<=N)

// output: T, MxK scores matrix 
// output: P, NxK loads matrix
// output: R, MxN residual matrix

cublasStatus status;

// maximum number of iterations 

int J = 10000;

// max error

double er = 1.0e-7; 

int n, j, k;
	
// transfer the host matrix X to device matrix dR 

double *dR = 0; 

status = cublasAlloc(M*N, sizeof(dR[0]), (void**)&dR);

if(status != CUBLAS_STATUS_SUCCESS)
 {
 fprintf (stderr, "! device memory allocation error (dR)\n"); 
 return EXIT_FAILURE;
 }    

status = cublasSetMatrix(M, N, sizeof(R[0]), R, M, dR, M);

if(status != CUBLAS_STATUS_SUCCESS) 
 {
 fprintf (stderr, "! device access error (write dR)\n"); 
 return EXIT_FAILURE;
 }

// allocate device memory for T, P

double *dT = 0; 

status = cublasAlloc(M*K, sizeof(dT[0]), (void**)&dT);
if(status != CUBLAS_STATUS_SUCCESS)
 {
 fprintf (stderr, "! device memory allocation error (dT)\n"); 
 return EXIT_FAILURE;
 }    

double *dP = 0; 

status = cublasAlloc(N*K, sizeof(dP[0]), (void**)&dP);

if(status != CUBLAS_STATUS_SUCCESS)
 {
 fprintf (stderr, "! device memory allocation error (dP)\n"); 
 return EXIT_FAILURE;
 }    
 
// allocate memory for eigenvalues

double *L; 

L = (double*)malloc(K * sizeof(L[0]));;

if(L == 0)
 {
 fprintf (stderr, "! host memory allocation error: T\n");  
 return EXIT_FAILURE;
 }
 
// mean center the data 

double *dU = 0; 
 
status = cublasAlloc(M, sizeof(dU[0]), (void**)&dU);
if(status != CUBLAS_STATUS_SUCCESS) 
 {
 fprintf (stderr, "! device memory allocation error (dU)\n"); 
 return EXIT_FAILURE;
 }     

cublasDcopy(M, &dR[0], 1, dU, 1);   

for(n=1; n<N; n++) 
 {
 cublasDaxpy (M, 1.0, &dR[n*M], 1, dU, 1);
 }

for(n=0; n<N; n++) 
 {
 cublasDaxpy (M, -1.0/N, dU, 1, &dR[n*M], 1);	
 }

// GS-PCA

double a; 
	
for(k=0; k<K; k++)
 {
 cublasDcopy (M, &dR[k*M], 1, &dT[k*M], 1);

 a = 0.0; 

 for(j=0; j<J; j++)
  {
  cublasDgemv ('t', M, N, 1.0, dR, M, &dT[k*M], 1, 0.0, &dP[k*N], 1);
  
  if(k>0)
   {
   cublasDgemv ('t', N, k, 1.0, dP, N, &dP[k*N], 1, 0.0, dU, 1);
   
   cublasDgemv ('n', N, k, -1.0, dP, N, dU, 1, 1.0, &dP[k*N], 1);
   }		
  cublasDscal (N, 1.0/cublasDnrm2(N, &dP[k*N], 1), &dP[k*N], 1);

  cublasDgemv ('n', M, N, 1.0, dR, M, &dP[k*N], 1, 0.0, &dT[k*M], 1);

  if(k>0)
   {
   cublasDgemv ('t', M, k, 1.0, dT, M, &dT[k*M], 1, 0.0, dU, 1);

   cublasDgemv ('n', M, k, -1.0, dT, M, dU, 1, 1.0, &dT[k*M], 1);
   }		

  L[k] = cublasDnrm2(M, &dT[k*M], 1);

  cublasDscal(M, 1.0/L[k], &dT[k*M], 1);			
 
  if(fabs(a - L[k]) < er*L[k]) break;

  a = L[k];

  }
 cublasDger (M, N, - L[k], &dT[k*M], 1, &dP[k*N], 1, dR, M);
 }
	
for(k=0; k<K; k++)
 {
 cublasDscal(M, L[k], &dT[k*M], 1);
 }
	
// transfer device dT to host T

cublasGetMatrix (M, K, sizeof(dT[0]), dT, M, T, M);

// transfer device dP to host P

cublasGetMatrix (N, K, sizeof(dP[0]), dP, N, P, N);

// transfer device dR to host R

cublasGetMatrix (M, N, sizeof(dR[0]), dR, M, R, M);
	
// clean up memory 

free(L); 

status = cublasFree(dP);

status = cublasFree(dT);

status = cublasFree(dR);

return EXIT_SUCCESS;
}


int print_results(int M, int N, int K, 
                  double *X, double *T, double *P, 
                  double *R)
{
int m, n, k;
	
// If M < 13 print the results on screen

if(M > 12) return EXIT_SUCCESS;

printf("\nX\n");

for(m=0; m<M; m++)
 {
 for(n=0; n<N; n++)
  {
  printf("%+f  ", X[id( m, n,M)]);
  }
 printf("\n");
 }

printf("\nT\n");

for(m=0; m<M; m++)
 {
 for(n=0; n<K; n++)
  {
  printf("%+f  ", T[id(m, n, M)]);
  }
 printf("\n");
 }

double a;

printf("\nT' * T\n");

for(m = 0; m<K; m++)
 {
 for(n=0; n<K; n++)
  {
  a=0; 
  for(k=0; k<M; k++) 
   {
   a = a + T[id(k, m, M)] * T[id(k, n, M)];
   }
  printf("%+f  ", a);
  }
 printf("\n");
 }		

printf("\nP\n");

for(m=0; m<N; m++)
 {
 for(n=0; n<K; n++)
  {
  printf("%+f  ", P[id(m, n, N)]);
  }
 printf("\n");
 }

printf("\nP' * P\n");

for(m = 0; m<K; m++)
 {
 for(n=0; n<K; n++)
  {
  a=0; for(k=0; k<N; k++) a = a + P[id(k, m, N)] * P[id(k, n, N)];
  printf("%+f  ", a);
  }
 printf("\n");
 }

printf("\nR\n");

for(m=0; m<M; m++)
 {
 for(n=0; n<N; n++)
  {
  printf("%+f  ", R[id( m, n,M)]);
  }
 printf("\n");
 }

return EXIT_SUCCESS;
}

\end{verbatim}

\end{document}